\documentclass{article}
\usepackage{graphicx}
\usepackage{epsfig}
\usepackage{amsmath}
\usepackage{amssymb}
\usepackage{amsfonts}
\usepackage{amsthm}
\DeclareMathAlphabet{\bi}{OML}{cmm}{b}{it}

\begin{document}

\title{Tensorial mobilities for accurate solution of transport problems in
models with diffuse interfaces}

\author{Matteo Nicoli, Mathis Plapp, Herv\'e Henry \\
Physique de la Mati\`ere Condens\'ee\\
\'Ecole Polytechnique, CNRS, 91128 Palaiseau, France}

\date{\today}

\maketitle

\begin{abstract}
The general problem of two-phase transport in phase-field models
is analyzed: the flux of a conserved quantity is driven by the
gradient of a potential through a medium that consists of domains
of two distinct phases which are separated by diffuse interfaces. 
It is shown that the finite thickness of the interfaces induces
two effects that are not present in the analogous sharp-interface 
problem: a surface excess current and a potential jump at 
the interfaces. It is shown that both effects can be 
eliminated simultaneously only if the coefficient of 
proportionality between flux and potential gradient (mobility) 
is allowed to become a tensor in the interfaces. This opens 
the possibility for precise and efficient simulations
of transport problems with finite interface thickness.
\end{abstract}


\section{Introduction}

Phase-field models  have recently enjoyed a rapidly growing popularity 
as a compact and elegant simulation tool for moving boundary problems
in such diverse fields as solidification \cite{Boettinger02,Steinbach09},
fluid dynamics \cite{Anderson98} or solid-state transformations 
\cite{Chen02,Wang10}. Their technical advantage resides in the implicit 
representation of interfaces by one or several phase fields, i.e.
fields that are defined in the entire space, take constant values within 
the bulk of each domain, and exhibit smooth but steep variations in
well-localized interfacial regions. 
The tedious procedure of front tracking is avoided  by introducing 
equations of motion for the phase fields that are coupled  
to the relevant transport variables.   
The price to pay for this advantage is the introduction of a new length 
scale into the model: the thickness $W$ of the phase-field front. For a given 
macroscopic problem, simulations with a phase-field model 
yield in general results that depend on the value of $W$.

In the field of crystal growth, great progress towards efficient and
precise simulations has been made by reducing this dependence on the 
interface thickness \cite{Karma98,Almgren99,Karma01,Echebarria04,Folch05}. 
This was made possible by a detailed analysis of the model equations 
using the method of matched asymptotic expansions, which is a systematic 
procedure to calculate the effective boundary conditions ``seen'' by 
the macroscopic transport field. 
Since this analysis is carried out
within a perturbation approach, these boundary conditions are naturally
expressed as a power series in $W$. Within  the phase-field community the 
limit  $W\to 0$ is referred to as {\sl the sharp-interface limit}.
When the corrections due to the finite interface
thickness are taken into account for choosing the model parameters, 
the accuracy of the phase-field method can be drastically improved. 
This procedure, which has been called 
{\sl thin-interface limit} \cite{Karma98}, 
has so far been worked out only for a few specific physical systems. 

To be more precise, let us consider the problem of solidification, in
which the relevant transport process that limits the growth of the
crystal is the diffusion of heat and/or solute. Two cases are completely
solved: the symmetric model, where the diffusion coefficients in the
two phases are identical \cite{Karma98}, and the one-sided model, in 
which no diffusion takes place 
within the solid phase \cite{Karma01,Echebarria04}.
However, so far no method has been found to eliminate all thin-interface
effects in the case of arbitrary diffusion coefficients in the two 
phases, despite some recent progress \cite{Almgren99,Ohno09,Steinbach09}
(for a more detailed discussion, see \cite{Plapp11}).

As will be pointed out here, part of this problem arises from the fact 
that for a truly two-sided model (with different diffusivity in each phase) 
even the stationary transport problem
{\em without} interface motion exhibits thin-interface effects. This 
prevents a solution of the problem by the antitrapping approach, which
has been successful for the one-sided model \cite{Karma01,Echebarria04}
and for the two-sided model
with vanishing diffusion current in one phase \cite{Ohno09}.

In fact, such thin-interface effects are fairly general and arise
in a whole class of problems, namely, two-phase transport through a 
complex structure. Examples of relevant physical situations 
are the conduction of electrical current or heat through a two-phase 
material with different conductivities, the magnetic flux through a 
two-phase material with different susceptibilities, or fluid flow though a 
porous medium with variations in permeabilities. At first sight, the
advantages of using the phase-field method in 
these cases are less obvious 
since the interfaces do not move and therefore the problem of front
tracking does not arise. However, the geometrical representation of 
complex-shaped surfaces can be difficult even without interfacial motion,
especially in three dimensions. Additionally, the use of a stationary 
phase-field function makes it possible to prescribe a given boundary 
condition at the interface in a straightforward  manner  
\cite{Kockelkoren03,Fenton05}, and to impose arbitrary boundary 
conditions at the border of a physical domain of complex shape, 
see \cite{Bueno-Orovio06,Li09} and references therein. 

The problem of representing a complex interface through a diffuse boundary
gains additional relevance 
because the use of tomographic methods for structure determination 
becomes more and more widespread. In such methods,
the structure representation takes 
the form of a matrix consisting of discrete pixels (or voxels in 
three dimensions) that contain binary or intensity data indicating 
whether a point in space is ``filled'' or ``empty''. From these data, the 
``true'' structure (represented for example by discrete sharp surface 
elements) has to be reconstructed by image analysis 
techniques. The phase-field 
method is an easy and robust method to obtain a smoothened representation 
of such data \cite{Benes04,Kay09}. 
It could be interesting to use directly this smoothened structure 
for accurate calculation of transport processes instead of going
through the additional steps of determining the ``sharp'' surface geometry.

However, it will be shown below that thin-interface effects are also present in
the  ``simple'' problem of two-phase transport, even without interface
motion. Therefore, these effects must be quantified and if possible eliminated. 
As we will demonstrate below, two effects that depend on the interface 
thickness are present: transport along the surface, and an interface 
resistance. In the standard phase-field formulation, where the transport 
is described by a scalar coefficient whose value depends on the phase field, 
these two effects cannot be eliminated simultaneously. In contrast, if the 
transport coefficient is allowed to become a {\em tensor} inside the diffuse 
interfaces, there are enough degrees of freedom in the model to eliminate 
both effects.

\begin{figure}[t!]
\centerline{
\epsfig{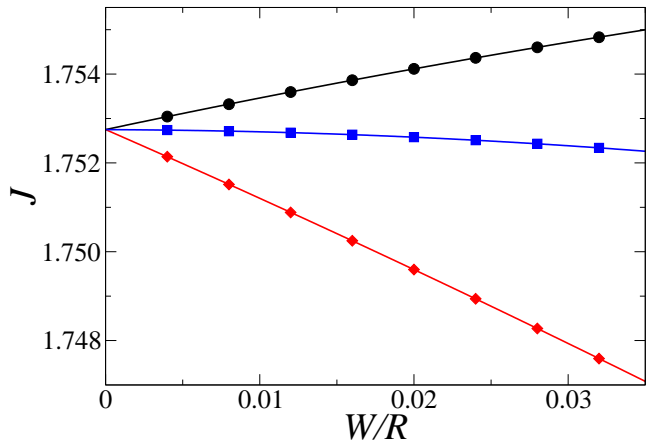}}
\caption{
\label{fig_1} 
(Color online) Total current as function of the ratio between the interface thickness
and the radius of the circle in the cases of  direct (black circles), inverse (red diamonds), and 
tensorial (blue squares) interpolation.
We solve the spherical inclusion problem with parameters: 
$M_1=1$, $M_2=1/2$, $R=0.25$, $h = 10^{-3}$, and $\rho = 10^{-6}$.
Lines are obtained by quadratic regression of the simulation data
(see the main text for further  details). The coefficients
of these regressions are contained in Tab.\ \ref{tab_1}. 
All units are arbitrary.}
\end{figure}

\section{Analysis}

\subsection{Problem formulation}

All the problems listed above have a common structure, namely, the flux 
of a conserved quantity is driven by a potential gradient,
\begin{equation}
\label{trans_1}
\bi{j} = - M(\phi) \nabla V,
\end{equation}
where $V$ is a potential.
The structure of Eq.\ \eqref{trans_1} is standard
in out-of-equilibrium thermodynamics: a linear relationship between
a flux and a thermodynamic driving force (the potential gradient).
The mobility coefficient $M(\phi)$ depends on the phase field $\phi$. 
If the transport process is electric conduction, $V$ is the electrostatic
potential and $M$ the conductivity; for diffusive mass transport
$V$ is the chemical potential and $M$ the atomic mobility.

The transported quantity satisfies a conservation law, which is
valid both in the bulk and at the surfaces, and for a time-independent 
solution (steady-state flow) reads
\begin{equation}
\nabla \cdot \bi{j} = 0.
\label{eq_Poisson}
\end{equation}
The problem specification is completed by a boundary condition for
the potential at the interfaces. We assume continuity of the
potential,
\begin{equation}
V_+ = V_-,
\label{continuity}
\end{equation}
where $V_+$ and $V_-$ are the values of the potential when the
interface is approached from the two sides. This corresponds 
to a rapid exchange of the transported quantity between the 
two sides of the interface.

Since we consider a fixed and immobile two-phase structure, the phase field 
is independent of time. We will assume that the two constant values that designate 
the two phases are $\phi=0$ and $\phi=1$, and that the field $\phi$ varies 
between these two limits continuously through a front region of width $W$. 
For the sake of concreteness, the reader may have in mind a sigmoid function 
such as $\left[1+\tanh(x/W)\right]/2$, but the explicit form of this function is not
important. The only hypothesis we make is that for a straight interface,
the profile of $\phi$ is odd with respect to the point $\phi=1/2$, that 
is, $\phi(x)=1-\phi(-x)$ for an interface centered at $x=0$. This is the
case in all standard phase-field models.
\begin{table}
\begin{center}
\begin{tabular}{|c|c|rr|c|}
\hline
interpolation type & $J_0$ & $c_1$ && $c_2$ \\[3pt]
\hline
direct   & 1.752751 & $0.73889  \times$& $10^{-1}$ & -0.277364 \\[2pt]
 inverse   & 1.752751 & $-1.51884 \times$&$ 10^{-1}$& -0.289096 \\[2pt]
 tensorial   & 1.752749 & $-1.352 \times$&$ 10^{-3}$ & -0.358637 \\[2pt]
\hline
\end{tabular}
\caption{\label{tab_1} Coefficients of the second order regression \mbox{$J(\epsilon) = J_0 + c_1 \epsilon + c_2 \epsilon^2$} 
for the three different interpolation methods. These coefficients are obtained from a quadratic regression to the data indicated by the symbols of Fig.\ \ref{fig_1}.}
\end{center}
\end{table}

\subsection{Surface current}

For simplicity of exposition, it is useful to focus on a concrete example. 
Consider the conduction of electric current through a two-phase material. 
Then, $V$ is the electrostatic potential, and $M(\phi)$ is the phase-dependent 
electric conductivity. Furthermore, consider a straight interface normal 
to the $x$ direction, centered at $x=0$. A potential gradient along the 
$y$ direction (along the interface) is imposed by sandwiching the material 
between two parallel plates located at $\pm L/2$ that are held at constant 
potentials $\pm U$. Since the phase field $\phi$ and hence the conductivity 
$M$ are constant along any line of constant $x$ (although their values 
differ for different values of $x$), Eq. (\ref{eq_Poisson}) yields a 
constant potential gradient $U/L$ directed along the $y$ direction. 
Therefore, the total current $J$ that flows between the two plates is 
given by
\begin{equation}
J=\int_{-\infty}^{\infty} M(\phi) \dfrac UL \; dx.
\end{equation}
Since we have considered a sample that extends to infinity, this current 
is clearly infinite. However, we will be concerned only with the excess
of this current with respect to the sharp-interface value. The latter is 
obtained as the current that would flow if $\phi(x)$ was a step function, 
that is, the space between the two plates is filled with material 1 of 
conductivity $M_1$ for $x<0$, and with material 2 of conductivity $M_2$ 
for $x>0$. This yields
\begin{equation}
\bar{J}=\int_{-\infty}^0 M_1 \dfrac UL \; dx + \int_0^\infty M_2 \dfrac UL \; dx.
\end{equation}
The difference between these two expressions is the excess of current 
$\delta J$
due to the variation of conductivity over a zone of finite thickness.
This excess
\begin{equation}
\delta J =\int_{-\infty}^0 \hskip -7pt \left[ M(\phi)-M_1\right] \dfrac UL \; dx + 
	\int_0^\infty \hskip -7pt \left[M(\phi)-M_2\right] \dfrac UL \; dx,
\end{equation}
is localized in the interface, and can therefore be interpreted as a 
additional current along the surface. It can be written as the product
of the potential gradient and a {\em surface conductivity}
\begin{equation}
M_s= \int_{-\infty}^0\hskip -2pt \left[M(\phi)-M_1\right] \; dx + \int_0^\infty \hskip -2pt \left[ M(\phi)-M_2\right] \; dx.
\label{ms}
\end{equation}
This surface transport coefficient has two obvious properties: 
(i) for an interface profile of fixed functional form, $\phi(x)=f(x/W)$, 
$M_s$ is proportional to the interface thickness (as can be shown by 
a simple change of variables), and (ii) it is strictly zero for any 
value of $W$ if 
\begin{equation}
\label{surf_cond}
\int_0^{-\infty} \left[M(\phi)-M_1\right] \; dx = \int_0^\infty \left[ M(\phi)-M_2\right] \; dx.
\end{equation}
In this case, the excesses of current on the two sides of the interface 
exactly compensate. For a phase-field profile that satisfies 
$\phi(-x)=1-\phi(x)$, this can be simply achieved by choosing
\begin{equation}
M(\phi)=M_1\phi + M_2(1-\phi).
\label{eq_parint}
\end{equation}
This will be called {\sl direct interpolation} in the following.
\subsection{Surface resistance}

Let us now analyze again a planar interface normal to the $x$ direction, 
but this time crossed by a steady current $J_\perp$ along $x$. In this case, 
the continuity equation immediately yields that the current is constant 
(independent of $x$). Then, the potential $V$ satisfies the simple equation
\begin{equation}
-M(\phi)\partial_x V = J_\perp.
\end{equation}
Integration along $x$ yields
\begin{equation}
V(x)-\overline V=-\int_0^x \frac {J_\perp}{M(\phi)} \; dx,
\label{eq_Vdiff}
\end{equation}
where $\overline V$ is the potential at $x=0$ (an integration constant). In 
contrast, if the interface was sharp, the potential would simply 
be given by
\begin{equation}
V_0(x)-\overline V= - \frac{xJ_\perp}{M_{1,2}}
\label{eq_Vsharp}
\end{equation}
for $x>0$ and $x<0$, respectively. Of course, outside the diffuse 
interfaces, the slopes of $V(x)$ are identical in Eqs. (\ref{eq_Vdiff})
and (\ref{eq_Vsharp}). Therefore, the asymptote of the diffuse-interface 
expression is of the form
\begin{equation}
V(x)-\overline V\approx - \frac{xJ_\perp}{M_{1,2}}+V_{+,-}
\end{equation}
for $x\to\pm\infty$. The constants $V_+$ and $V_-$ (the interface
potentials ``seen'' from the region outside the diffuse interface) 
are readily obtained from the matching of this expression to 
Eq. (\ref{eq_Vdiff}),
\begin{equation}
V_{+,-}=\overline V + J_\perp\int_0^{\infty,-\infty} \left[\frac {1}{M(\phi)}-\frac{1}{M_{1,2}}\right]\; dx.
\end{equation}
Of particular interest is the fact that these surface potentials can 
be different, in contradiction to the assumption of Eq. (\ref{continuity}). 
The difference $\delta V=V_+-V_-$ can be written as the 
product of the current $J_\perp$ and an interface resistance
\begin{equation}
R_s=\int_{-\infty}^0 \left[\frac{1}{M(\phi)}-\frac{1}{M_1}\right] \; dx + 
\int_0^\infty \left[\frac{1}{M(\phi)}-\frac{1}{M_2}\right] \; dx.
\end{equation}
This resistance is often called Kapitza resistance and has been
frequently observed in experiments and simulations 
\cite{Kapitza41,Wolf83,Swartz89,Barrat03,Xue03}.
Again, it is obvious that it is proportional to the interface 
thickness, and that it vanishes if the integral is exactly 
zero. This can be achieved for any value of $W$ by the interpolation
\begin{equation}
\frac {1}{M(\phi)} = \frac{1}{M_1}\phi + \frac{1}{M_2}(1-\phi).
\label{eq_perpint}
\end{equation}
This will be called {\sl inverse interpolation} in the following.

\begin{figure}[t!]
\centerline{
\epsfig{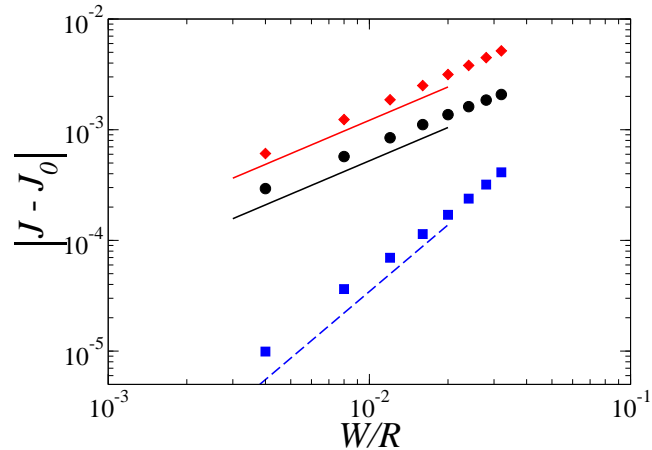}}
\caption{
\label{fig_2} 
(Color online) Rate of convergence of the three interpolations as function of the ratio
between the interface thickness and the radius of the disk. 
Black circles, red diamonds, and blue squares stand for direct [Eq.\ \eqref{eq_parint}], inverse [Eq.\ \eqref{eq_perpint}], 
and tensorial interpolation [Eq.\ \eqref{eq_tensint}],
respectively. The physical and numerical parameters are the same used in Fig.\ \ref{fig_1} and $J_0 = 1.75275$. 
Red and black solid lines are guide to eye with slope equal to $1.0$, while blue dashed line has slope equal to $2.0$.
All units are arbitrary.}
\end{figure}

\subsection{Tensorial mobility}

In summary, the two interface effects (surface current and surface 
resistance) can each be eliminated by a specific choice for the 
interpolation function of the mobility. Since these 
interpolations are mutually exclusive, it seems as if necessarily 
one of the two effects must remain nonzero. However, the current 
is a vector quantity, and the two effects are linked to distinct 
components of the current vector: the excess surface conductivity 
is relevant only for the components parallel to an interface, 
whereas the surface resistance modifies the boundary conditions for 
the normal component. 
Inside the two phases, where each medium is isotropic, the Curie principle 
requires to choose a scalar mobility to relate the current and the 
potential gradient. However, in the presence of an interface, isotropy 
of space is broken and a tensorial transport coefficient is permitted. 
Indeed, the gradient of the phase field can be readily used to define 
the interface normal $\bi{n} = \nabla\phi / |\nabla\phi|$,
which provides a second direction that is independent of the potential 
gradient. Then, we can define the transport coefficient by
\begin{equation}
\label{eq_tensint}
\bi{M}(\phi)=M_\perp \bi{n} \otimes \bi{n} + M_{||} (\mathbf {1}-\bi{n} \otimes \bi{n}),
\end{equation}
where $\mathbf{1}$ is the unit tensor, with two independent interpolation 
functions $M_\perp(\phi)$ and $M_{||}(\phi)$. If we interpolate $M_\perp$ 
according to Eq. (\ref{eq_perpint}) and $M_{||}$ according to 
Eq. (\ref{eq_parint}), both thin-interface effects are eliminated.
Hence, for this  interpolation the transport problem defined by 
Eq.\ \eqref{trans_1}  becomes
\begin{equation}
\label{trans_2}
\bi{j} = -\bi{M}(\phi)\cdot \nabla V,
\end{equation}
with components (in two dimensions)
\begin{eqnarray}
j_x &=& M_{xx} \partial_x V + M_{xy} \partial_y V, \\[5pt]
j_y &=& M_{xy} \partial_x V + M_{yy} \partial_y V.
\end {eqnarray}
Here, we have designated by $M_{ij}$  the elements of the symmetric tensor $\bi{M}(\phi)$. 
The simple calculations developed above are valid only for planar
interfaces. However, for a sufficiently smooth interface (that is,
with a local radius of curvature $R$ satisfying $R\gg W$), 
a local curvilinear coordinate system can be defined in which the 
above relations remain valid at least up to second order in $\epsilon= W/R$.
It should be mentioned that a similar strategy has been used recently to 
develop efficient phase-field models for surface diffusion \cite{Gugenberger08}.

\begin{figure}[t!]
\centerline{
\epsfig{file=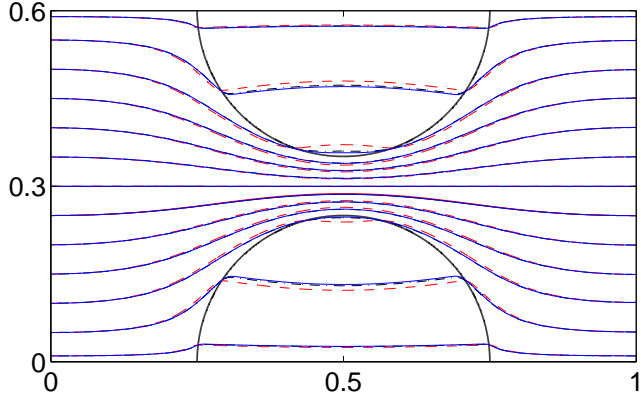,width=0.7\textwidth,clip=}}
\caption{
\label{fig_4} 
(Color online) Streamlines for the second geometrical configuration studied. 
The black dash-dotted, red dashed, and blue solid lines
correspond to direct, inverse, and tensorial interpolation, respectively. 
The simulation parameters are 
$M_1=1$, $M_2=0.1$, $R=0.25$, $h = 10^{-3}$, $\rho = 10^{-6}$, and $W=4h$.
For a larger version of this picture see \cite{epaps}.
All units are arbitrary.}
\end{figure}

\section{Numerical validation}

We quantify the thin-interface effects in the three different interpolations of
the transport coefficient by solving the  problem defined by
Eq.\ \eqref{trans_1} [or \eqref{trans_2}] and Eq.\ \eqref{eq_Poisson} 
in a simple geometrical setup. We consider a square domain 
${\cal D}\equiv\{(x,y)\in [0,1]\times[0,1]\}$ with Dirichelet boundary 
conditions on the lateral edges and  zero flux at the upper and bottom edges. 
More precisely, we impose \mbox{$V(0,y) = 1$}, \mbox{$V(1,y) = -1$}, and
\mbox{$\partial_y V(x,0) = \partial_y V(x,1) = 0$}.
In the center of the domain we place a disk of radius $R$. The mobility 
coefficient takes a value of $M_1$ ($M_2$) outside (inside) the disk,
respectively.  
We refer to this geometrical setup as spherical inclusion problem.
 
By combining the flux equation Eq.\ \eqref{trans_1} [or  its tensorial counterpart \eqref{trans_2}] 
with the conservation law of our problem, i.e.\  Eq.\ \eqref{eq_Poisson}, we obtain
an elliptic equation for the case of direct and inverse interpolation
\begin{equation}
\nabla\cdot \left[ M(\phi) \, \nabla V \right] = 0,
\label{elliptic_1}
\end{equation}
whereas the tensorial interpolation leads to the following equation
\begin{equation}
\label{elliptic_2}
 \nabla\cdot \left[ \bi{M}(\phi) \cdot \nabla V \right] = 0.
\end{equation}
Details about the discretization and the method used to solve these
equations can be found in the appendix. 
 
\begin{figure}[t!]
\centerline{
\epsfig{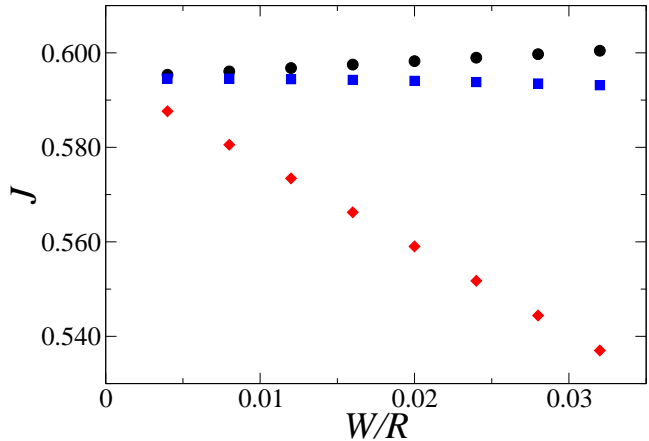}}
\caption{
\label{fig_3} 
(Color online) Total current as function of the ratio between the interface thickness
and the radius of the circle for  direct (black circles), inverse (red diamonds), and 
tensorial (blue squares) interpolations. The simulation parameters are
the same of Fig.\ \ref{fig_4}. 
All units are arbitrary.}
\end{figure} 
 
The thin-interface effects 
arising in  the spherical inclusion problem are quantified by measuring  
the total flux at $x = 1$,
\begin{equation}
J =  \int_0^1 j_x(1,y)\, dy 
\end{equation}
and by plotting $J$ as a function of the ratio $\epsilon=W/R$ between the 
interface width $W$ and the radius of the disk $R$, see Fig.\ \ref{fig_1}.
As shown in Fig.\ \ref{fig_1}, the three interpolations converge to 
the same value of $J_0$ when $\epsilon \to 0$. The estimation of $J_0$ 
is given by a quadratic regression in $\epsilon$ of the simulation data.
The coefficients of the regressions are listed in Table\ \ref{tab_1}. 
Within the truncation error [that is ${\rm O}\left( h^2\right) \sim 10^{-6}$] 
the three interpolations give the same value of $J_0 = 1.752750\pm 10^{-6}$, 
as shown in the second column of Table\ \ref{tab_1}.

In addition, we have estimated the rate of convergence of $J(\epsilon)$ to the 
sharp-interface limit $J_0$ for the three interpolations of the mobility. 
As shown in Fig.\ \ref{fig_2}, the direct and inverse 
interpolations converge only linearly with $\epsilon$ to this limiting value,
whereas the tensorial interpolation suppress linear thin-interface effects,
which leads to a convergence that is almost quadratic in $\epsilon$. 

For further illustration, we consider a second geometrical configuration 
for\-med by two half-disks of radius $R$ placed in a rectangular domain with 
the same boundary condition of the spherical inclusion problem, see 
Fig.\ \ref{fig_4}. We choose a small value of the ratio between $M$ inside 
and outside the half-disks, e.g.\ $M_2/M_1 = 0.1$, and in Fig.\ \ref{fig_4} 
we plot the streamlines for this configuration. In this configuration, the
flux is constricted in the narrow space between the two half disks. The 
difference between the three interpolations is largest for streamlines which 
are locally almost tangent to the half-disks; note the divergence of the 
different lines close to the tips of the half-disks. As shown by 
Eq.\ \eqref{surf_cond}, choosing $M$ according to the direct interpolation 
cancels exactly the surface conductivity $M_s$ for a flat interface.
Therefore, thin-interface errors affecting streamlines which are parallel 
to the boundary of variation of the  transport coefficient are more pronounced 
in case of inverse interpolation (red dashed lines) than in case of direct  
interpolation (black dash-dotted lines), see Fig.\ \ref{fig_4}.
In Fig.\ \ref{fig_3}, we show again the values of the total current as
a function of the ratio $\epsilon$, and again the tensorial interpolation 
performs much better than the other two, as expected.

\section{Conclusion}
We have investigated a phase-field model with a mobility tensor, in 
which normal and parallel components of the flux are interpolated 
with distinct functions of the phase field. Contrary to phase-field 
models with scalar mobilities, this method makes
it possible to eliminate at the same time the additional surface 
diffusion and the surface resistance, which are both linked to the
finite thickness of the interfaces. This opens the possibility to
perform accurate simulations of two-phase transport problems with
enlarged interface thickness, which can lead to dramatic savings 
in computation time.

The complete elimination of thin-interface effects to first order
in the interface thickness is also a prerequisite for the development
of a quantitative crystal growth model for arbitrary ratio of the
diffusion coefficients in the two phases. Indeed, a tensorial diffusion 
coefficient can remove both surface diffusion and the Kapitza 
resistance \cite{Plapp11} from such models. 
Even if a second order asymptotic analysis of a time dependent version of the
tensorial problem \eqref{eq_tensint} has been recently performed \cite{Ngoc}, 
this methodology is still not capable of removing all thin-interface effects 
from phase-field models with a non-stationary $\phi$.
The obstacle that needs to be overcome for the successful development 
of such a model is to find a coupling of the transport equation
to an evolution equation for the phase field that yields the
correct boundary conditions for the transport field at {\em moving} 
interfaces. We hope to be able to report on this problem in the near future.

\section*{Acknowledgements}
Support provided by the European Commission
through the MODIFY (FP7-NMP-2008-SMALL-2, Code
228320) research project is greatly acknowledged.
\mbox{M.\ N.\ }also acknowledge partial support by 
MICINN (Spain) Grant No.\ FIS2009-12964-C05-01.

\appendix

\section*{Numerical Methods}

The differential operators involved in Eq.\ \eqref{elliptic_1}  
and \eqref{elliptic_2}  are approximated by finite differences. The  
domain $\cal D$ is discretized by an uniform square mesh with 
elements of area $h^2$. The fields $M$, $V$, $j_x$, and $j_y$  
are placed in different grids according to the standard marker and 
cell (MAC) method \cite{Pozrikidis_book}. In the primary grid we 
locate $\phi$, $M$ ($M_{xx}$ and $M_{yy}$ in case of tensorial interpolation), 
and $V$ 
while in the two staggered grids (one for each direction) we put 
each component of the flux $\bi{j}$. The primary nodes $(i,j)$ have 
coordinates 
\begin{eqnarray}
x &=& h(i+1/2), \\[5pt]
y &=& h(j+1/2), 
\end{eqnarray}
whereas the two  
staggered grids are shifted of $h/2$ in the $x$ ($y$) direction 
for $j_x$ ($j_y$), respectively. In what follows, with the 
notation $[\dots]_{i,j}$ we mean that the field inside square 
brackets is evaluated at the position $(i,j)$ with respect to the 
primary grid. For example, at the nodes of the $x$ staggered 
grid we have $j_{x\, i,j}$ that is the value of $x$ component 
of the flux between the nodes $(i,j)$ and $(i+1,j)$ of the 
primary grid, i.e.\ $\left[Êj_x\right]_{i+1/2,j}$.
  
In order to compute $\bi{j}$ with this MAC arrangement we have to specify the values of $M$ at the nodes of the staggered grids.
 This is readily  done by using the averaging operator on the $k$ coordinate  $\mu^\pm_k$, that is
 \begin{eqnarray}
 &&\left[M\right]_{i\pm1/2,j} = \mu^\pm_x M_{i,j} = \left( M_{i\pm1,j} + M_{i,j}\right)/2, \\[5pt]
 &&\left[M\right]_{i,j\pm 1/2} = \mu^\pm_y M_{i,j} = \left( M_{i,j\pm1} + M_{i,j}\right)/2.
 \end{eqnarray}
 Hence, the two components of the flux $\bi{j}$ read
 \begin{eqnarray}
&& j_{x\, i,j} = \left[ j_x\right]_{i+1/2,j} = \mu^+_x M_{i,j} \, \triangle^+_x V_{i,j}  / h,\\[5pt]
&&  j_{y\, i,j}  = \left[j_y\right]_{i,j+1/2} = \mu^+_y M_{i,j} \, \triangle^+_y V_{i,j}  / h,
\end{eqnarray}
where  $h^{-1} \triangle^+_k$ is the standard forward difference operator acting on the $k$ direction. 
Finally, Eq.\ \eqref{elliptic_1} is discretized by employing the backward difference operators $h^{-1}\triangle^-_k$
\begin{equation}
\label{delliptic_1}
\dfrac{1}{h^2}\sum_{k=x,y} \triangle^-_k\left(\mu^+_k M_{i,j} \, \triangle^+_k V_{i,j} \right) = 0.
\end{equation}

Evidently, the discretization of the elliptic problem arising from the tensorial interpolation $\bi{M}$  is 
more involved. In fact, two parts of Eq.\ \eqref{elliptic_2} mix $x$ and $y$ derivatives, i.e.\
$\partial_x \left( M_{xy} \partial_y V \right)$ and $\partial_y \left( M_{xy} \partial_x V \right)$. 
In order to guarantee second order accuracy and 
maximum  compactness of the discrete stencil it is convenient to place the $M_{xy}$ component of the tensor $\bi{M}$ on a third
grid, whose  nodes are shifted by $h/2$ in the two directions \cite{Gerritsma96}.  
The tensorial interpolation produces an additional contribution to $j_x$ 
\begin{equation}
\left[M_{xy} \partial_y V\right]_{i+1/2,j} = \mu_y^-\left( M_{xy\, i,j}\mu_x^+ \triangle^+_y V_{i,j}\right) /h,
\end{equation}
and to  $j_y$
\begin{equation}
\left[M_{xy} \partial_x V\right]_{i,j+1/2} = \mu_x^-\left( M_{xy\, i,j}\mu_y^+ \triangle^+_x V_{i,j}\right) /h.
\end{equation}
As before, by applying  backward differentiation  we obtain the discrete version of Eq.\ \eqref{elliptic_2}
\begin{equation}
\label{delliptic_2}
\begin{split}
\frac{1}{h^2} & \Big\{ \triangle^-_x\left[\mu^+_x M_{xx\, i,j} \, \triangle^+_x V_{i,j} + 
	\mu_y^-\left( M_{xy\, i,j}\mu_x^+ \triangle^+_y V_{i,j}\right) \right] + \\
& \triangle^-_y \left[ \mu_x^-\left( M_{xy\, i,j}\mu_y^+ \triangle^+_x V_{i,j}\right)+ \mu^+_y M_{yy\, i,j} \, \triangle^+_y V_{i,j}  \right] 
	 \Big\} = 0.
\end{split}
\end{equation}

The MAC arrangement ensures a second order accuracy  of the truncation error of  the two elliptic problems~\cite{Pozrikidis_book,Gerritsma96}.
Eq.\ \eqref{delliptic_1} and \eqref{delliptic_2} are two linear systems of equations where the unknowns are  the values of the
potential $V_{i,j}$. These linear systems can be easily solved through any 
iterative method, for example by the  Successive Over Relaxation (SOR) \cite{NumericalRecipes}.
To find an accurate solution of these problems we iterate the SOR algorithm until the maximum residue 
at the nodes of the primary grid
\begin{equation}
\rho = \max_{i,j} \big|\left[\nabla \cdot \bi{j} \right]_{i,j}\big|,
\end{equation}  
is comparable with the truncation error, i.e.\ $\rho \leq h^2$.


\end{document}